\documentclass[twocolumn,aps,prl,floats,floatfix,superscriptaddress,showpacs]{revtex4}

\usepackage{graphicx}
\begin{document}

%
%Now come the authors with their affiliation
%
\author{A.D. Christianson}
\affiliation{Oak Ridge National Laboratory, Oak Ridge, TN 37831 USA}
\author{M.D. Lumsden}
\affiliation{Oak Ridge National Laboratory, Oak Ridge, TN 37831 USA}
\author{M. Angst}
\affiliation{Oak Ridge National Laboratory, Oak Ridge, TN 37831 USA}
\author{Z. Yamani}
\affiliation{National Research Council, Canadian Neutron Beam Center,
Chalk River, Ontario, Canada K0J 1J0}
\author{W. Tian}
\affiliation{Oak Ridge National Laboratory, Oak Ridge, TN 37831 USA}
\affiliation{Ames Laboratory, Iowa State University, Ames, IA 50011 USA}
\author{R. Jin}
\affiliation{Oak Ridge National Laboratory, Oak Ridge, TN 37831 USA}
\author{E.A. Payzant}
\affiliation{Oak Ridge National Laboratory, Oak Ridge, TN 37831 USA}
\author{S.E. Nagler}
\affiliation{Oak Ridge National Laboratory, Oak Ridge, TN 37831 USA}
\author{B.C. Sales}
\affiliation{Oak Ridge National Laboratory, Oak Ridge, TN 37831 USA}
\author{D. Mandrus}
\affiliation{Oak Ridge National Laboratory, Oak Ridge, TN 37831 USA}

\title{Three Dimensional Magnetic Correlations in Multiferroic LuFe$_2$O$_4$}
\date{\today }
\begin{abstract}
We present single crystal neutron diffraction measurements on multiferroic LuFe$_2$O$_4$.  Magnetic reflections are observed below transitions at 240 and 175 K indicating that the magnetic interactions in LuFe$_2$O$_4$ are 3-dimensional (3D) in character.  The magnetic structure is refined as a ferrimagnetic spin configuration below the 240 K transition.  Below 175 K a significant broadening of the magnetic peaks is observed along with the build up of a diffuse component to the magnetic scattering.

\end{abstract}
\pacs{: 77.84.-s, 75.30.Kz, 28.20.Cz, 25.40.Dn}
\maketitle

Materials that offer the possibility of simultaneously controlling magnetic and electric degrees of freedom are the subject of intense interest \cite{Cheong}.  Recently, multiferroic materials have been identified that show large coupling between electric and magnetic degrees of freedom.  Ferroelectricity driven by either magnetic or charge ordering appears to be the origin of the large coupling and, hence, understanding the underlying electronic interactions is crucial for further insight into multiferroicity \cite{Cheong}.

LuFe$_2$O$_4$ has attracted attention as a novel ferroelectric material where ferroelectricity is driven by the electronic process of charge ordering of Fe$^{2+}$ and Fe$^{3+}$ ions and for indications of % a concomitant
coupling between electronic and magnetic degrees of freedom \cite{Ikeda,Subramanian,Xiang,Zhang,Nagano}.  LuFe$_2$O$_4$ is a member of the $R$Fe$_2$O$_4$ ($R =$ rare earth element) family, the physical properties of which depend strongly on oxygen stoichiometry.  For example, nearly stoichiometric YFe$_2$O$_4$ exhibits three-dimensional (3D) magnetic order while oxygen deficient YFe$_2$O$_4$ exhibits two-dimensional (2D) magnetic order \cite{Funahashi}.  LuFe$_2$O$_4$ exhibits multiple phase transitions.  2D charge correlations are observed below 500 K, while below ~320 K 3D charge order is established, roughly coinciding with the onset of ferroelectricity \cite{Ikeda,Yamada}.  Magnetic order appears below 240 K and 2D ferrimagnetic order has been suggested by neutron scattering studies \cite{Iida}.  However, strong sample dependent behavior observed in other members of $R$Fe$_2$O$_4$ \cite{Funahashi} suggests that unraveling the interesting behavior of LuFe$_2$O$_4$ requires paying due attention to sample quality.
	
In this letter we present extensive neutron diffraction measurements from 20 to 300 K on high quality single crystals of LuFe$_2$O$_4$.  We report several new findings that provide information about the underlying magnetic interactions.  First, our measurements indicate that below 240 K 3D magnetic correlations exist with magnetic intensity appearing at (1/3 1/3 L) where L may take on integer \emph{and} 1/2-integer values.  The magnetic structure is refined with a ferrimagnetic spin configuration with a propagation vector of (1/3 1/3 0).  The magnetic intensity appearing on peaks where L is a 1/2-integer is a consequence of the charge ordering at $\sim$320 K.  In addition, evidence is presented for a second transition at 175 K with significant changes in magnetic peak intensities and broadening of many reflections.	
	
Single crystals of LuFe$_2$O$_4$ were grown by floating-zone-melting, using an oxygen partial pressure tuned by a CO/CO$_2$ mixture to control oxygen stoichiometry \cite{Iida2}. For CO/CO$_2$ ratio close to 2.7 the magnetic behavior exhibits two sharp magnetic transitions in contrast to previous single crystal magnetization measurements \cite{Iida} which show only a single transition.  The magnetic behavior is qualitatively similar to that of stoichiometric YFe$_2$O$_4$ where previous work with slightly varying oxygen concentration showed that stoichiometric samples are characterized by much sharper magnetic transitions and reduced residual low temperature ($T$) susceptibility \cite{Inazumi}. We take the sharpness of the observed transitions and reduced residual low temperature susceptibility, in light of previous measurements on YFe$_2$O$_4$, as strong evidence that these LuFe$_2$O$_4$ crystals have nearly ideal oxygen stoichiometry and are extremely homogeneous \cite{Stoich}.  Two crystals of the growth batch in which the sharpest transitions were observed, denoted S1 and S2, were selected for neutron diffraction measurements.  S2 is slightly more homogeneous as judged from magnetization curves.

\begin{figure}
\centering
\includegraphics[width=\columnwidth,clip]{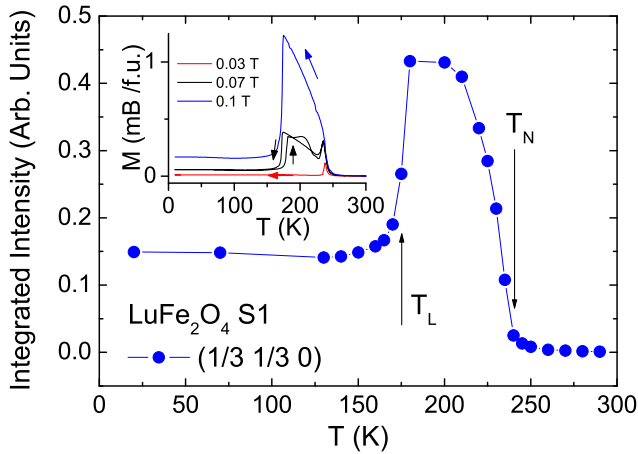}
\caption{(Color online) Integrated Intensity for the (1/3 1/3 0) magnetic peak measured in the (HK0) plane.  The error bars are smaller than the point size.  The inset shows field-cooled, c-axis magnetization data at 0.03 (red), 0.07 (black), and 0.1 Tesla (blue).  Arrows indicate whether the measurement was made upon warming or cooling.}
\label{tdep}
\end{figure}

Neutron diffraction measurements were performed on the N5 triple-axis spectrometer (TAS) at the Canadian Neutron Beam Center at Chalk River Laboratories and the HB1 TAS at the High Flux Isotope Reactor at Oak Ridge National Laboratory.  Neutrons with incident energies of 14.56 meV (N5) and 30.5 meV (HB1) were selected and horizontal collimations of 30'-36'-16.4'-66' (N5)  and 48'-40'-40'-70' (HB1) were used.  Pyrolytic graphite (PG) (002) was used for monochromator and analyzer and PG filters were placed in the scattered beam to suppress higher order contamination.
	
Figure 1 shows the integrated intensity vs. $T$ for the (1/3 1/3 0) magnetic peak providing evidence for two phase transitions, one at 240 K ($T_N$) and another at 175 K ($T_L$) \cite{Kakurai}.  The presence of two transitions is corroborated by the c-axis magnetization (inset Fig. 1) of a crystal from the same batch as the neutron scattering samples.  The sharpness of $T_L$ in both neutron and magnetization data suggests that $T_L$ is of 1st order, which is confirmed by hysteresis in the magnetization.
	
\begin{figure}
\centering
\includegraphics[width=\columnwidth,clip]{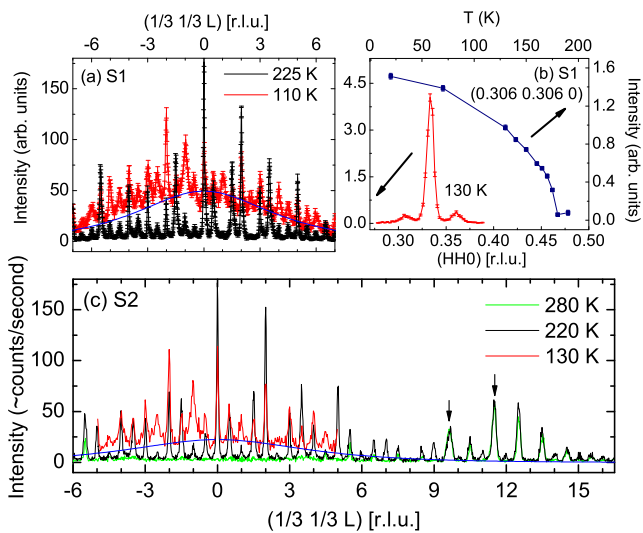}
\caption{(Color online) (a) and (c) show scans along (1/3 1/3 L) for S1 and S2.  The solid blue lines are described in the text.  The error bars in (c) have been omitted for clarity but are comparable to those displayed in (a).  The arrows indicate peaks contaminated by Aluminum background scattering.  (b) Displays data showing the appearance below $T_L$ of a new set of satellites indexed by (1/3$\pm \delta$ 1/3$\pm \delta$ 3L/2) with $\delta \sim$0.027.}
\label{scans}
\end{figure}

Figure 2 (a) and (c) displays scans along (1/3 1/3 L) at several temperatures.  At 280 K peaks at large values of L are readily visible with 1/2-integer indices (Fig. 2(c)).  As pointed out previously \cite{Ikeda}, these peaks are consistent with a $\sqrt{3} \times \sqrt{3} \times$ 2 unit cell containing 36 Fe atoms.  Measurements at larger values of momentum transfer, Q,  indicate that the peaks are not magnetic and, in accord with previous work, they are attributed to the onset of 3D charge order at 320 K \cite{Yamada}.
	
Cooling below 240 K, new intensity appears at integer values of L.  This intensity is strongest at small L and for L $>$ 16 it has diminished to the point where it is difficult to discern from the background.  Measurements at (2/3 2/3 L) and (4/3 4/3 L) reinforce the conclusion that the intensity on integer L positions only occurs for sufficiently small values of Q.  Similarly,  the intensity at 1/2-integer positions become enhanced at small Q, but not at larger Q where the intensity is essentially unchanged from that observed at 280 K.  Such Q-dependence is expected for scattering from the magnetic moments of Fe$^{2+}$ and Fe$^{3+}$ indicating that peaks indexed by integer values of L originate principally from an ordered magnetic sublattice.  This also indicates that the changes of the L 1/2-integer peaks are predominantly due to magnetic order rather than modifications of the charge ordering configuration.  Thus, the neutron scattering data demonstrates 3D magnetic correlations below 240 K.  However, we do note that the magnetic reflections are not resolution limited along L suggesting a finite correlation length.
	
To allow for quantitative comparison with models for the spin configuration, a large number of reflections were measured at 220 K by scanning along the (1/3 1/3 L), (2/3 2/3 L), and (4/3 4/3 L) directions.  To solve for the magnetic structure, representational analysis was performed to consider those magnetic structures which are symmetry allowed from the parent $R\bar{3}m$ space group \cite{Wills,BasIreps}.  This analysis assumes that the onset of charge order at 320 K does not significantly affect the symmetry-allowed magnetic order. Initially, a magnetic propagation vector of (1/3 1/3 1/2) was considered and symmetry analysis yielded two possible irreducible representations.  However, both possibilities yielded L 1/2-integer reflections several orders of magnitude more intense than the L-integer reflections.  This is in contrast to observation where the 1/2-integer peaks are comparable to or weaker than the integer reflections (see Fig. 2 (c)).  Consequently, we concluded that the magnetic structure is described by the ordering wavevector (1/3 1/3 0), the presence of 1/2-integer reflections occurring as a result of the charge ordering which decorates the lattice with differing magnetic moment on Fe$^{2+}$ and Fe$^{3+}$ sites with a periodicity of (1/3 1/3 1/2).	 Representational analysis with the (1/3 1/3 0) wavevector again yielded two allowed irreducible representations.  For spins pointing along the c-axis, as suggested by the magnetization measurements, these representations correspond to ferromagnetic (FM) or antiferromagnetic (AFM) alignment of the two spins of the primitive basis (see Fig. 3).  The AFM case can be ruled out immediately as the magnetic structure, including symmetry equivalent wavevectors, does not yield intensity at the (1/3 1/3 0) position.  The FM coupling between spins in the basis, the only remaining symmetry allowed possibility, results in a ferrimagnetic structure as shown in Fig. 3 for the symmetry equivalent propagation vectors (1/3 1/3 0), (-2/3 1/3 0), and (1/3 -2/3 0).
	
\begin{figure}
\centering
\includegraphics[width=\columnwidth,clip]{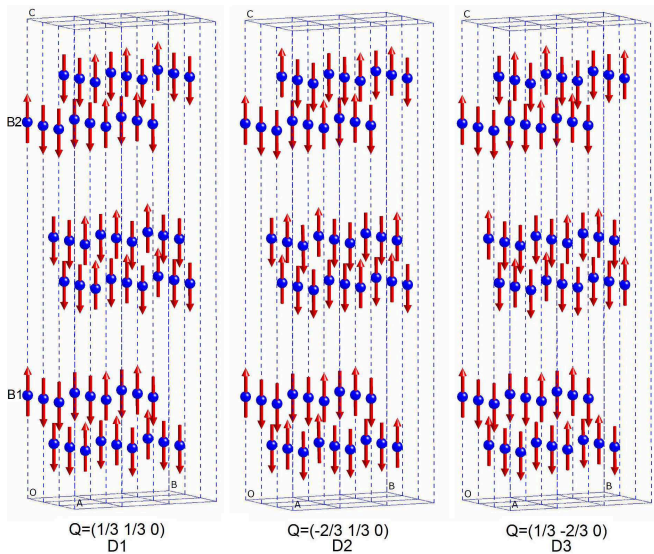}
\caption{(Color online) Magnetic structure of LuFe$_2$O$_4$ at $T$=220 K.  The three magnetic domains correspond to the three symmetry equivalent magnetic propagation vectors (1/3 1/3 0), (1/3 -2/3 0), and (-2/3 1/3 0) labeled as D1, D2, and D3 respectively.  B1 and B2 denote the primitive basis ((0, 0, 0.22) and (0, 0, 0.78)).}
\label{structure}
\end{figure}

\begin{figure}
\centering
\includegraphics[width=\columnwidth,clip]{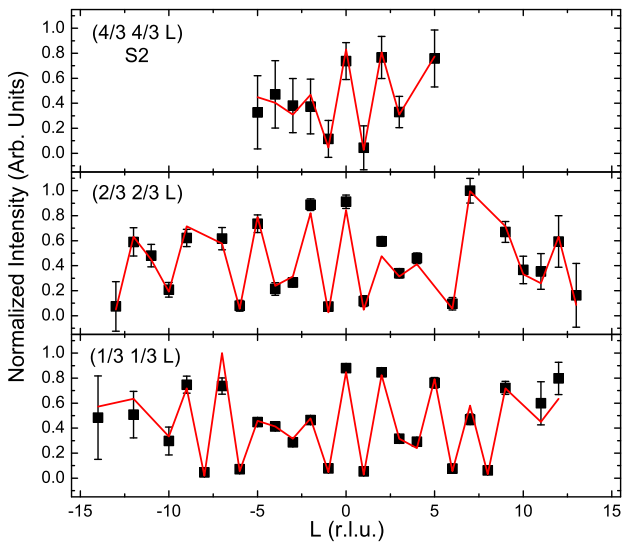}
\caption{(Color online) Peak intensity for scans along (1/3 1/3 L), (2/3 2/3 L) and (4/3 4/3 L) corrected as described in the text.  The solid red line represents the result of the model calculation described in the text. }
\label{model}
\end{figure}

Determining the agreement between the model and experiment was complicated by reflections that were not resolution limited along the c-axis.  To account for this, the TAS resolution function was fully simulated \cite{Reslib} and the data from several reflections were fit assuming a Lorentzian lineshape along the c-axis with resolution limited peaks in the hexagonal basal plane.  This nicely reproduced the lineshape and yielded a Lorentzian HWHM ($\Delta$) of 0.0257 r.l.u. corresponding to a correlation length ( $\xi=1/\Delta$) of 156~\AA.  The resolution correction was then obtained by convolving this lineshape with the resolution function for all measured reflections.  For simplicity, the magnetic moments were fixed to be the same on all sites.  This assumption is justified given the charge order which occurs at 320 K and the domain population of this charge ordered state.  To accurately reproduce the 1/2-integer L peaks, the correct charge ordering scheme would need to be included, but it is not necessary to describe the integer-L reflections.  Therefore we proceeded to fit 58 observed integer-L reflections with a model using only 4 parameters, two domain population factors, an overall scale factor, and a Debye-Waller factor.  The model agrees very well with the data (reduced  $\chi^{2}=$1.39).
	
To extract meaningful domain populations, it was recognized that the Lorentzian width varied slightly between the three domains yielding correlation lengths of 155~\AA, 130~\AA, and 160~\AA~ for domains D1, D2, and D3 (see Fig. 3).  The resolution correction was modified to account for this and the resulting data corrected for resolution, magnetic form factor, and spin polarization factor are plotted in Fig. 4 along with the calculated intensities.  The $\chi^{2}$ was improved slightly to 1.37 and the resulting domain population ratios are 0.85:0.57:1.  Comparable domain population ratios are observed for S1, a sample taken from the same growth.
Whether or not the domain populations are an intrinsic property of LuFe$_2$O$_4$ or are sample dependent is, at present, unclear.
Domain D1 yields a symmetric pattern in L with peaks described by L=3n while D2 (D3) produce peaks at L=3n+1 (3n-1) for H=1/3 and 4/3, the order reversed for H=2/3.  The ratio of D2:D3 (0.57:1) is responsible for the asymmetry in the diffraction pattern for $\pm$L.  The proposed magnetic structure shown in Figure 3 is ferrimagnetic with an excess of 1/3 of the spins pointing along the c-axis. This results in a saturation ordered moment of 1/3 $\times$ 4.5 $\mu_B$/Fe $\times$ 2 Fe/FU or 3  $\mu_B$/FU.  This value is very close to the saturation magnetization in higher fields ($H$ $>$ 2 T) in our crystals (not shown) as well as to the value reported in Ref. \cite{Iida} further corroborating the magnetic structure.
The smaller magnetic moment observed in the magnetization at $H$ $<$ 2 T is a consequence of the formation of domains and will be discussed in more detail elsewhere.

Below $T_L$ an additional component to the scattering builds up which is extremely broad along (1/3 1/3 L) but sharp along (HH0) (Fig 2(a) and (c)).  This diffuse scattering appears to be magnetic in origin as emphasized by the solid blue lines in Fig. 2 (a) and (c) which are proportional to the product of the form factor squared, a Debye-Waller factor, and the polarization factor (1+$\widehat{Q}_z^{2}$).  The diffuse scattering is stronger in S1 which, as judged by magnetization measurements, is slightly less homogenous than S2 suggesting that the presence of 2D magnetic short range order is a property of samples that are not sufficiently homogeneous.  This bears some similarity to YFe$_2$O$_4$ where the dimensionality of the magnetic interactions (2D or 3D) depends on the oxygen stoichiometry \cite{Funahashi}.  Figure 2 also shows that below $T_L$ profound changes occur in the magnetic peaks along (1/3 1/3 L).  The intensity along (1/3 1/3 L) for magnetic reflections changes rather dramatically with some peaks becoming more intense (e.g. (1/3 1/3 1)) and some peaks becoming less intense (e.g. (1/3 1/3 0)).  Thus 3D magnetic correlations persist below $T_L$ albeit with a shorter correlation length than found for $T_N$.  Finally, we note that scans along (110) have revealed the existence of a new set of satellite peaks of unknown origin indexed as (1/3$\pm \delta$ 1/3$\pm \delta$ 3L/2) where $\delta \sim$0.027 (see figure 2 (b)) below $T_L$.
	
The increase in linewidth on many peaks below $T_L$ is consistent with the introduction of stacking faults at a structural phase transition \cite{Hendricks-Teller}.  Evidence that $T_L$ involves a structural component is provided by an extinction related increase in intensity of strong structural Bragg peaks as well as broadening of those peaks below 175 K.  CuK$\alpha$ powder x-ray diffraction as well as the single crystal neutron data shows a contraction (expansion) of the $a$ ($c$) lattice constant from 350 K to 100 K but does not reveal any sign of a structural distortion.  Stoichiometric YFe$_2$O$_4$ provides an example of a related system which exhibits phase transitions with a structural component at similar temperatures to the two transitions found in LuFe$_2$O$_4$ below 300 K \cite{Funahashi}.
%To prove and detail a structural transition occurring at $T_L$ careful, high-resolution crystallographic work is required.

Although the linewidth of the peaks below $T_L$ and the broad diffuse component to the scattering makes a full solution of the low $T$ magnetic structure very difficult, some general conclusions are possible.  The 3D magnetic correlations in LuFe$_2$O$_4$ are intrinsically sensitive to disruptions along the c-axis as the superexchange path between Fe-O bilayers must pass through not one but two oxygen ions and thus the driving force for the changes in the magnetic structure below $T_L$ may be related to the introduction of stacking faults as discussed above.  The stacking arrangement would result in local magnetic order which deviates from the ferrimagnetic state and may mix in states which are locally AFM.  If we consider the symmetry allowed AFM state, the most intense reflection is the (1/3 1/3 $\pm$1) reflection, the most strongly enhanced and broadened reflection seen experimentally in the low $T$ state.  Additionally, the (1/3 1/3 0) peak is absent in the AFM structure, qualitatively consistent with the observed strong reduction in the intensity of this reflection seen at low temperatures. Furthermore, local AFM order would result in decreased c-axis magnetization, leading naturally to the similar temperature dependence of the magnetization and the (1/3 1/3 0) intensity (Fig. 1). A detailed quantitative analysis of the neutron scattering and magnetization data below 175 K, beyond the scope of the present work, should shed further light on the complex magnetic behavior in LuFe$_2$O$_4$.

In conclusion, we show that that LuFe$_2$O$_4$ has two transitions below 300 K.  Both transitions involve a 3D magnetically correlated structure with a finite correlation length along the c-axis.  Whether the correlation length is an intrinsic property or is the result of disorder, most likely oxygen stoichiometry, is yet to be elucidated.  Below $T_N$ a ferrimagnetic spin configuration is found with a magnetic propagation vector of (1/3 1/3 0) with magnetic intensity occurring at (1/3 1/3 L) where L is 1/2-integer arising due to the charge ordering at 320 K. Theoretical models taking into account the 3D nature of the magnetic interactions as well as the sequence of magnetic phase transitions described above should provide insight into the multiferroic behavior of LuFe$_2$O$_4$.

We acknowledge useful discussions with V.O. Garlea.  ORNL is managed by UT-Battelle, for the DOE under Contract No. DE-AC05-00OR22725.


\begin{thebibliography}{}

\bibitem{Cheong} S.-W. Cheong, and M. Mostovoy, Nature Mat. \textbf{6}, 13 (2007).

\bibitem{Ikeda} N. Ikeda, \emph{et al.}, Nature \textbf{436}, 1136 (2005).

\bibitem{Subramanian} M. A. Subramanian, \emph{et al.}, Adv. Mater. \textbf{18}, 1737 (2006).

\bibitem{Xiang} H. J. Xiang and M. -H. Whangbo, Phys. Rev. Lett. \textbf{98}, 246403 (2007).

\bibitem{Zhang} Y. Zhang, \emph{et al.}, Phys. Rev. Lett. \textbf{98}, 247602 (2007).

\bibitem{Nagano} A. Nagano, \emph{et al.}, Phys. Rev. Lett. \textbf{99}, 217202 (2007).

\bibitem{Funahashi} S. Funahashi, \emph{et al.}, J. Phys. Jpn. \textbf{53}, 2688 (1984).

\bibitem{Yamada} Y. Yamada, \emph{et al.}, Phys. Rev. B \textbf{62}, 12167 (2000).

\bibitem{Iida} J. Iida, \emph{et al.}, J. Phys. Soc. Jpn. \textbf{62}, 1723 (1993).

\bibitem{Iida2} J. Iida, S. Takekawa, and N. Kimizuka, J. Crystal Growth \textbf{102}, 398 (1990).

\bibitem{Inazumi} M. Inazumi, \emph{et al.}, J. Phys. Soc. Jpn. \textbf{50}, 438 (1981).

\bibitem{Stoich}
%The window of observation the magnetic behavior of truly stoichiometric of LuFe2O4 is expected to be narrower than it is in YFe2O4.  For the present samples, TGA can comfirm stoichiometry to better than 1\%.
%The window of observation of the magnetic behavior of truly stoichiometric LuFe$_2$O$_4$ is likely narrower than in YFe$_2$O$_4$, explaining the non-observation of these properties in prior studies.
The non-observation of this magnetic behavior in LuFe$_2$O$_4$ in prior studies is likely due to a smaller stoichiometry range in which it occurs, compared to YFe$_2$O$_4$.  With the limited amount of single crystal material available, standard methods for direct determination of oxygen content are too inaccurate to quantitatively characterize the extremely sensitive stoichiometry dependence.


\bibitem{Kakurai} During the course of this work we became aware of unpublished work by K. Kakurai, et al., that gives indications of similar behavior.

\bibitem{Wills} A. S. Wills,  Physica B \textbf{276}, 680 (2000), http://www.chem.ucl.ac.uk/people/wills/.

\bibitem{BasIreps} J. Rodrguez-Carvajal, BasIreps, \\ http://www.ill.fr/dif/Soft/fp/php/downloads.html.

\bibitem{Reslib} A. Zheludev, Reslib, \\ http://neutron.ornl.gov/$\sim$zhelud/reslib/.

\bibitem{Hendricks-Teller} For example, S. Hendricks and E. Teller, J. Chem. Phys. \textbf{10}, 147 (1942).


\end{thebibliography}
\end{document}